\documentclass[aps,prl,twocolumn,showpacs,a4paper,superscriptaddress,amsmath,amssymb]{revtex4}

\usepackage[dvips]{graphicx}
\usepackage[dvips]{color} 
\usepackage{subfigure}

\renewcommand{\vec}[1]{\mathbf #1}

\newcommand{\diff}[2]{\frac{\mathrm{d} #1}{\mathrm{d} #2}}

\newcommand{\pdiffdiff}[2]{\frac{\partial^2 #1}{\partial #2^2}}
\renewcommand{\d}{\mathrm d}

\renewcommand{\hbar}{\hslash}

\begin{document}
 
\title{Proximity effect-assisted absorption of spin currents in
  superconductors}
 
\author{Jan Petter Morten} \affiliation{Department of Physics,
  Norwegian University of Science and Technology, NO-7491 Trondheim,
  Norway} \affiliation{Centre for Advanced Study at the Norwegian
  Academy of Science and Letters, Drammensveien 78, NO-0271 Oslo,
  Norway} 
\author{Arne Brataas} \affiliation{Department of Physics,
  Norwegian University of Science and Technology, NO-7491 Trondheim,
  Norway} 
\affiliation{Centre for Advanced Study at the Norwegian
  Academy of Science and Letters, Drammensveien 78, NO-0271 Oslo,
  Norway} 
\author{Gerrit E. W. Bauer} \affiliation{Kavli Institute of
  NanoScience, Delft University of Technology, 2628 CJ Delft, The
  Netherlands} \affiliation{Centre for Advanced Study at the Norwegian
  Academy of Science and Letters, Drammensveien 78, NO-0271 Oslo,
  Norway} 
\author{Wolfgang Belzig} \affiliation{University of
  Konstanz, Department of Physics, D-78457 Konstanz, Germany}
\affiliation{Centre for Advanced Study at the Norwegian Academy of
  Science and Letters, Drammensveien 78, NO-0271 Oslo, Norway}
\author{Yaroslav Tserkovnyak} \affiliation{Department of Physics and
  Astronomy, University of California, Los Angeles, California 90095,
  USA} \affiliation{Centre for Advanced Study at the Norwegian Academy
  of Science and Letters, Drammensveien 78, NO-0271 Oslo, Norway}

\date{December 17, 2007}
 
\begin{abstract}
  The injection of pure spin current into superconductors by the
  dynamics of a ferromagnetic contact is studied theoretically. Taking
  into account suppression of the order parameter at the interfaces
  (inverse proximity effect) and the energy-dependence of spin-flip
  scattering, we determine the temperature-dependent ferromagnetic
  resonance linewidth broadening. Our results agree with recent
  experiments in Nb$\vert$permalloy bilayers [C.  Bell \textit{et
    al.}, arXiv:cond-mat/0702461].
\end{abstract}
 


 
\maketitle

Cooper pairs in conventional superconductors are spin-singlet states
and therefore cannot carry a spin current. Some aspects of the
resilience of the superconducting state against spin-current injection
have been experimentally demonstrated in hybrid
ferromagnet-superconductor spin valves \cite{Gu:prb02-short}, switches
\cite{s-switch-misc}, and $\pi$-junctions \cite{pi-misc}. In these
experiments, the spin current flow in the superconducting state can
only be inferred via charge current measurements. This complicates the
understanding of the spin current flow in superconductors.

Injection of a pure spin current into a superconductor has recently
been demonstrated by Bell \textit{et al.} \cite{bell:cond-mat/0702461}
in ferromagnet$\vert$superconductor structures under ferromagnetic
resonance (FMR) conditions, in which the precessing magnetization acts
as a ``spin pump'' \cite{Fpump-theory}. The spin angular momentum lost
by the ferromagnet can be observed directly in terms of an increased
broadening of the FMR spectrum.  In this Letter we demonstrate
theoretically that the spin transport thus measured as a function of
temperature and device/material parameters offers direct insight into
spin-flip relaxation and the inverse proximity effect in
superconductors. Our theory agrees well with the recent experimental
results \cite{bell:cond-mat/0702461}, and we provide suggestions and
predictions for future experiments.

The theoretical challenge of spin-pumping into superconductors as
compared to normal conductors is the strong energy dependence of
quasiparticle transport properties around the superconducting energy
gap \cite{Morten:Sspindiffusion}.  Also, the energy dependent
spin-flip scattering rates caused by spin-orbit coupling or magnetic
impurities differ. Experiments that directly probe spin transport,
such as Ref. \onlinecite{bell:cond-mat/0702461}, therefore provide
unique information about the spin-flip scattering mechanism. A
complicating factor is the inverse proximity effect
\cite{Sillanpaa:epl01} that suppresses the superconducting order
parameter close to a metallic interface with ferromagnets like Ni, Co,
and Fe. The resulting spatial dependence of the superconducting gap
requires solution of the full transport equations in the entire
superconducting layer. The spin currents measured at such interfaces
therefore serve as probes of superconducting correlations in magnetic
heterostructures, and the temperature dependence of the FMR linewidth
near and below the critical temperature can provide a wealth of
information about spin-flip processes and superconducting proximity
physics, with potential implications for different areas of mesoscopic
physics.

 \begin{figure}[h]
  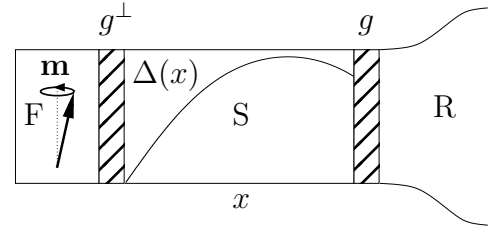
  \caption{The ferromagnet$\vert$superconductor$\vert$spin reservoir
    (F$\vert$S$\vert$R) structure. Precession of magnetization
    $\vec{m}(t)$ pumps spins into S, which can diffuse and dissipate
    in R. The F$\vert$S interface spin-mixing conductance for spins
    polarized transverse to the magnetization direction is $g^{\perp}$
    and the S$\vert$R interface conductance is $g$. The superimposed
    superconducting gap $\Delta(x)$ is suppressed close to the
    interfaces (inverse proximity effect).}
  \label{fig:fsn}
\end{figure}

In the following we develop a theory of energy-dependent spin pumping
at a ferromagnet$\vert$superconductor interface and the resulting
spectral spin current flow in the superconductor.  We consider a
diffusive metallic heterostructure consisting of a superconducting
layer (S) of thickness $L$ that is sandwiched by a ferromagnet (F) of
thickness $d$ and a spin reservoir (``spin sink'') (R), see
Fig. \ref{fig:fsn}. The slowly precessing magnetization $\vec{m}(t)$
emits a spin current that is transversely polarized with respect to
the instantaneous magnetization direction \cite{Fpump-theory}. The
spin current that flows through S is immediately dissipated upon
reaching R. R thus increases the sensitivity of the experiments to the
spin transport properties of S. R represents either a cap of an
efficient spin-flip scattering material such as Pt or a large
reservoir of a high mobility metal \cite{Fpump-theory}. We assume
sufficient thermal anchoring so that heating from absorbed FMR
microwave radiation can be disregarded.

The magnetization dynamics is determined by the generalized
Landau-Lifshitz-Gilbert equation,
\begin{align}
  \diff{\vec{m}}{t}=-\gamma\vec{m}\times\vec{H}_\text{eff}+\frac{G_0}{\gamma M_\text{S}}\vec{m}\times\diff{\vec{m}}{t}+\frac{\gamma}{M_\text{S}V}\vec{I}_\text{s}.
\label{eq:LLG}
\end{align}
Here $\gamma$ is the gyromagnetic ratio, $\vec{H}_\text{eff}$ is the
effective magnetic field, $M_\text{S}$ is the saturation
magnetization, and $V$ is the volume of the ferromagnet. The intrinsic
dissipation in the bulk ferromagnet is parametrized by the Gilbert
damping constant $G_0$. $\vec{I}_\text{s}$ is the total spin
(i.e. angular momentum) current generated by the precessing
ferromagnet. This loss of angular momentum is equivalent to an
interface contribution to the magnetization damping and is observable
in terms of the enhanced FMR linewidth broadening. Our task is to
evaluate the effect of superconducting correlations on
$\vec{I}_\text{s}$. The results can be summarized in terms of an
effective resistor model for the spin transport. We find an
energy-dependent spin transport resistance of S in series with the
spin-mixing resistance $r^\perp=1/g^{\perp}$ of the F$\vert$S
interface in the normal state and the conventional resistance $r=1/g$
of the S$\vert$R interface.

To illustrate the physics we first sketch the results for $\vec{m}(t)$
rotating in the $xy$-plane and in the absence of spin-flip scattering
(the derivation for the general situation will be outlined
subsequently). The magnetization then emits a time-independent spin
current that is polarized along the $z$-axis \cite{Fpump-theory}.  The
superconducting condensate consists of spin-singlet Cooper pairs. A
spin current can therefore only be carried in S by excited
quasiparticles. Since the low-energy density of quasiparticle states
is suppressed by superconducting correlations, the spin transport
resistivity is enhanced when S is in the superconducting state,
resulting in reduced spin injection from the ferromagnet. The
energy-dependent spin resistance is governed by a spectral Ohm's law,
\begin{align}
  R_\text{eff}^{\perp}(E)=\frac{r^{\perp}}{N(0,E)}+\int_0^L d x'\frac{\rho_\text{L}(x',E)}{A}+\frac{r}{N(L,E)} \, ,
  \label{eq:Reffnospinflip}
\end{align}
where $\rho_\text{L}=1/(hN_0D_\text{L})$ is the effective resistivity
of the superconductor for spin transport in units of $e^2/h$, $N_0$ is
the density of states at the Fermi level in the normal state,
$D_\text{L}(x,E)$ and $N(x,E)$ are the effective spin diffusion
coefficient and the normalized density of state at position $x$ and
energy $E$, respectively \cite{Morten:Sspindiffusion}. At zero
temperature, the relevant quasiparticle energy $E$ is determined by
the FMR frequency which is typically $f_{\rm FMR} \sim 10~G$Hz. For
BCS superconductors $hf_{\rm FMR}/\Delta_0\approx
0.3~\text{K}/T_\text{c}$ where $\Delta_0$ is the bulk zero-temperature
energy gap and $T_\text{c}$ the critical temperature of the
superconductor. For small-angle precession, the effective ``rotation''
frequency can be introduced as $f\sim\phi f_{\rm FMR}$, where $\phi$
is the angle of precession. Thus the relevant energy scale for
FMR-generated excitations is in practice expected to be much smaller
than $hf_{\rm FMR}$, and the characteristic energy of pumped electrons
is set by the temperature, see Eq.~(\ref{eq:Gnospinflip}) below. At
the F$\vert$S interface $N(x=0,E)\approx 1$ due to the inverse
proximity effect (see below). $R_\text{eff}^\perp$ depends on
temperature through the local gap $\Delta(x,T)$ which determines
$N(x,E)$ and $\rho_\text{L}(x,E)$. The spin current loss of the
ferromagnet is consistent with the Gilbert phenomenology in terms of
an increased damping parameter $G$. It is determined by the spin
angular momentum escape rate through S and reads
\begin{align}
  G=G_0+\frac{(g_\text{L}\mu_\text{B})^2}{2\pi\hbar}\frac{1}{d}\int d E\frac{-\d f_\text{FD}(E)/\d E}{AR_\text{eff}^\perp(E)},\label{eq:Gnospinflip}
\end{align}
where $g_\text{L}$ is the $g$-factor, $\mu_\text{B}$ is the Bohr
magneton, $A$ is the sample cross section area, and $f_\text{FD}$
is the Fermi-Dirac distribution function.
 
At temperatures $T\ll T_\text{c}$, $\Delta(x)$ as a function of the
distance from the F$\vert$S interface approaches the bulk value on the
scale of the bulk superconducting coherence length
$\xi_0=\sqrt{\hslash D/2\pi k_\text{B}T_\text{c}}$. Since the relevant spin resistivity
$\rho_\text{L}(x,E)$ and thus $R_\text{eff}^\perp$ are very large for
$E<\Delta$, $\xi_0$ sets the penetration length scale for spin current
into the superconductor. At low temperatures and $L>\xi_0$ the Gilbert
damping \eqref{eq:Gnospinflip} will therefore be weakly enhanced. On
the other hand, at $T\lesssim T_\text{c}$ the gap is suppressed
throughout S and transport channels at energies $E\gtrsim \Delta$
become accessible. $R_\text{eff}^\perp$ and the Gilbert damping then
approach the normal state values.

Spin-flip scattering in S dissipates spin current emitted from F, and
enhances $G$ by suppressing the back-flow of spins into the
ferromagnet. The spin-flip length in the normal state is given by
$l_\text{sf}=\sqrt{D\tau_\text{sf}}$, where $D$ is the normal state
diffusion coefficient. We take spin-flips into account that are caused
by magnetic impurities as well as spin-orbit coupling at impurities in
terms of the spin-flip rate
$1/\tau_\text{sf}=1/\tau_\text{m}+1/\tau_\text{so}$
\cite{Morten:Sspindiffusion}. The spin-orbit coupling respects the
symmetry of singlet Cooper pairs, whereas the pair-breaking scattering
by magnetic impurities suppresses superconductivity and reduces
$T_\text{c}$. Below $T_\text{c}$, the spin-flip rates in S depend on
energy. For $E<\Delta$ spin-flip rates both due to spin-orbit coupling
and magnetic impurities are suppressed. For $T\ll T_\text{c}$ and
$L>\xi_0$, the Gilbert damping will therefore be weakly enhanced. On
the other hand, for $E>\Delta$ the spin-flip rate due to magnetic
impurities is enhanced whereas the spin-flip rate due to spin-orbit
coupling is similar to that in the normal state. We therefore predict
a non-monotonic temperature dependence of the Gilbert damping close to
the critical temperature when spin-flip is dominated by magnetic
impurities. Experimental data indicate that $l_\text{sf}>\xi_0$ for
typical S.  $l_\text{sf}=48$ nm and $\xi_0=13$ nm has been reported
for Nb \cite{Gu:prb02-short} (which is used in
Ref. \onlinecite{bell:cond-mat/0702461}) whereas
$l_\text{sf}=1.1~\mu$m and $\xi_0=124$ nm for Al
\cite{Jedema:prb03,Boogaard:prb04}. When $L\leq \xi_0$ spin-flip in S
is therefore inefficient since $L\leq \xi_0 < l_\text{sf}$ in these
materials. We are then allowed to disregard spin-flip scattering
\cite{Fpump-theory}. On the other hand, when $L \gg l_\text{sf}$ the
spin current never reaches R so that $G$ is governed exclusively by
spin-flip in S for all temperatures. In the interesting regime where
$l_\text{sf}\approx L$, the full theoretical treatment sketched in the
following has to be invoked in order to compute the competing effects
that determine $G$.

The total spin current leaving the ferromagnet in the
F$\vert$S$\vert$R heterostructure can be expressed as an energy
integral over the balance of the spectral pumping and back-flow
currents $\vec{I}_\text{s}=\int\d
E(\vec{i}_\text{s}^\text{inj}-\vec{i}_\text{s}^\text{back})$. The spin
current injected into S by the precessing magnetization is
\cite{buttiker:133,Fpump-theory}:
\begin{align}
  \vec{i}_\text{s}^\text{inj}(E)=&\frac{\hbar N(0,E)}{4\pi}\frac{f_\text{FD}(E-hf/2)-f_\text{FD}(E+hf/2)}{hf}\nonumber\\
  &\times\left(g^{\perp}_\text{r}\vec{m}\times\diff{\vec{m}}{t} + g^{\perp}_\text{i}\diff{\vec{m}}{t}\right),
  \label{eq:spinpump}
\end{align}
where $f$ is the instantaneous rotation frequency. Here,
$g^\perp_\text{r}$ and $g^\perp_\text{i}$ are the real and imaginary
parts of spin-mixing conductance. For metallic interfaces,
$g^\perp_\text{r}\gg g^\perp_\text{i}$ \cite{brataas:physrep06}. We
therefore disregard the ``effective field'' $g^\perp_\text{i}$ in
\eqref{eq:spinpump}, although it contributes to the interface boundary
conditions discussed below. The magnetization damping that follows
from \eqref{eq:spinpump} is frequency dependent beyond the Gilbert
phenomenology. We have checked numerically that the $f$-dependent
terms contribute weakly to the damping even when $hf\lesssim\Delta_0$
for the parameters studied. We therefore restrict attention to the
linear response regime in which the Fermi-Dirac functions in
\eqref{eq:spinpump} can be expanded to first order in $hf$. This leads
to frequency-independent enhanced Gilbert damping in
\eqref{eq:LLG}. The spectral back-flow of spin current into F induced
by the spin accumulation on the S side is
\begin{align}
\vec{i}_\text{s}^\text{back}(E)=&-\frac{N(0,E)}{4\pi}g^{\perp}_\text{r}\vec{h}_\text{TS}(0,E).
  \label{eq:backflow}
\end{align}
The nonequilibrium spin distribution function $\vec{h}_\text{TS}(x,E)$
can be computed by Keldysh transport theory
\cite{Morten:Sspindiffusion}. In the S bulk, the total spin current
$\vec{I}_\text{s}(x)=\hbar A N_0 \int_{-\infty}^\infty d E D_\text{L}(E,x)\partial_x \vec{h}_\text{TS}(x,E)/2$
follows from the diffusion equation
\begin{align}
  \left(N\partial_t+\partial_x D_\text{L} \partial_x -\frac{\alpha^\text{m}_\text{TSTS}}{\tau_\text{m}}-\frac{\alpha^\text{so}_\text{TSTS}}{\tau_\text{so}}\right)\vec{h}_\text{TS}=0\label{eq:hTS}.
\end{align}
Diffusion through S is taken to be instantaneous on the scale of the
FMR frequency as long as $f<D/L^2$ and/or $f\ll 1/\tau_\text{sf}$ so
that $\vec{h}_\text{TS}$ in \eqref{eq:hTS} becomes time-independent.
$\alpha^\text{m(so)}_\text{TSTS}=[\text{Re}\,\cosh\theta]^2+(-)[\text{Re}\,\sinh\theta]^2$
are energy-dependent renormalization factors for the spin-flip rates
due to magnetic impurities (spin-orbit coupling), and the energy
dependent spin diffusion coefficient
$D_\text{L}/D=\alpha^\text{so}_\text{TSTS}$.  The spectral properties
of the superconductor parametrized by $\theta(x,E)$ are determined by
the Usadel equation for the retarded Green function
$\hat{G}^\text{R}=\hat{\tau}_3\cosh \theta+i\hat{\tau}_2\sinh \theta$,
\begin{align}
  \frac{\hbar D}{2} \pdiffdiff{\theta}{x}=i \Delta\cosh(\theta)-i
  E\sinh(\theta)+\frac{3}{8}\frac{\hbar}{\tau_\text{m}}\sinh(2\theta),\label{eq:retardedusadel}
\end{align} to be solved with the BCS gap equation
$\Delta=(N_0\lambda/2)\int_0^{E_\text{D}}d E
\tanh(E/2k_\text{B}T)\text{Re}\,\sinh(\theta)$
\cite{Morten:Sspindiffusion}. Here, $E_\text{D}$ is the Debye cut-off
energy and $\lambda$ the interaction parameter.

The boundary condition for the diffusion equation (\ref{eq:hTS}) is
conservation of spin current at the interfaces. At $x=0$, $\hbar
AN_0D_L \partial_x\vec{h}_{\text{TS}}/2=\vec{i}_{\text{s}}^\text{inj}-\vec{i}_{\text{s}}^\text{back}$.
We use boundary conditions derived in
Ref. \onlinecite{kupriyanov:1988} for \eqref{eq:retardedusadel} at the
S$\vert$R interface. At the F$\vert$S interface we impose complete
suppression of superconducting correlations, $\theta(x=0,E)=0$ for the
following reasons. The large exchange energy in transition metal
ferromagnets completely suppress superconducting correlations, so that
the F adjacent to S is a source of incoherent particles. Additionally,
spin dependent interface scattering at the S side
\cite{huertas-hernando:epjb05} induces an effective pair-breaking
exchange field, which we estimate as $B_\text{eff}=\hbar
g^{\perp}_\text{i}/e^2g_\text{L}\mu_\text{B}N_0A\xi_0$
\cite{bauer:prl126601-short}. Here, $N_0A\xi_0$ is the number of
states at the Fermi energy within $\xi_0$ from the interface. With
$g^{\perp}_\text{i}\approx 0.05g_\text{Sh}$, where $g_\text{Sh}$ is
the Sharvin conductance \cite{brataas:physrep06}, and approximating
$N_0$ by the free-electron value, $\mu_\text{B}B_\text{eff}$ is
comparable to $\Delta_0$, e.g,
$\mu_\text{B}B_\text{eff}(\text{Nb})\sim
0.56\,$meV$,\,\mu_\text{B}B_\text{eff}(\text{Al})\sim 69\,\mu$eV.  The
bulk F exchange splitting and the induced $B_\text{eff}$ by
spin-dependent interface scattering leads to a vanishing gap (and
$\theta$) at the F$\vert$S interface \cite{Sarma:63,Belzig:prb00}.

The spin diffusion equation \eqref{eq:hTS} can be solved analytically
in the absence of spin-flip, proving \eqref{eq:Reffnospinflip}. We now
use the full machinery sketched above to make contact with
experimental results for a F$\vert$S device (without R) similar to
sample C in Ref. \onlinecite{bell:cond-mat/0702461}. Numerically
computing $\vec{I}_\text{s}$ including spin-flip caused by magnetic
impurities \cite{poli:cond-mat/0707.2879-short}, we obtain the
enhanced Gilbert damping $G$ from \eqref{eq:LLG}. In the experiment, F
is a permalloy layer with $d=2$~nm, and $g_\text{L}=2.1$. S is Nb with
$L=70$~nm, bulk critical temperature $T_\text{c0}=8.91$~K,
$l_\text{sf}=48$~nm, and $D=5.41$~cm$^2$s$^{-1}$
\cite{koperdraad:9026,Gu:prb02-short}. For the interface conductances
we use $Ar=3$~f$\Omega$m$^2$ \cite{Bass:jmmm99}. We find
$G-G_0=0.777\times 10^8\,\text{s}^{-1}$ at $T_\text{c}/2=3.6$~K and
$1.19\times 10^8\,\text{s}^{-1}$ in the normal state.  When the
inhomogeneous linewidth broadening is small, the width of the FMR
spectra are proportional to $G$ and the experimental data gives
$[G(T>T_\text{c})-G(T=T_\text{c}/2)]/G(T>T_\text{c})\approx
21~\%$. Using $G_0=0.7\times 10^8~\text{s}^{-1}$ \cite{Fpump-theory}
we obtain $22~\%$. The measured reduction of the Gilbert damping upon
cooling the sample from above $T_\text{c}$ to $T_\text{c}/2$ agrees
\emph{quantitatively} with our calculation.

\begin{figure}
  \includegraphics[scale=0.666]{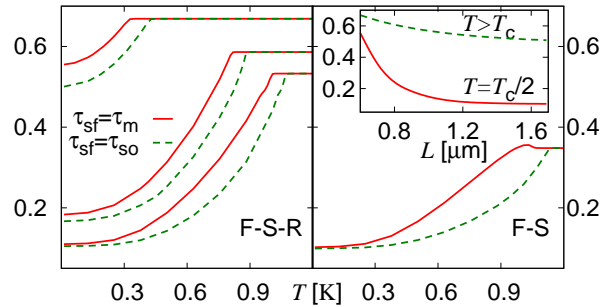}
  \caption{\label{fig:gilbertsf}Calculated $G-G_0$ [10$^8$ s$^{-1}$]
    (same ordinate in all plots). Red solid (green dashed) lines for
    system where $\tau_\text{sf}=\tau_\text{m}$
    ($\tau_\text{sf}=\tau_\text{so}$). Left panel: F$\vert$S$\vert$R
    system with $L$ [nm] from top to bottom: 600, 900, 1300. Right
    panel: F$\vert$S system (no R) with $L=900$ nm. Inset: $L$
    dependence [$\mu$m] of $G-G_0$ for $T> T_\text{c}$ (green dashed
    line) and $T\ll T_\text{c}$ (red solid line).}
\end{figure}

We can make additional predictions for the Gilbert damping in
F$\vert$S$\vert$R systems, focusing on Al as S since its spin-flip
length is much larger than that of Nb, and as a weak coupling
superconductor is better described by BCS theory. The Al material
parameters are $T_\text{c0}=1.26$ K, $l_\text{sf}=1.1~\mu$m, and
$D=160$ cm$^2$s$^{-1}$. In the left panel of Fig. \ref{fig:gilbertsf}
we show the temperature dependence of $G-G_0$ for three different
thicknesses $L$ when spin-flip is induced exclusively by either
magnetic disorder or spin-orbit coupling to impurities. In contrast to
spin-orbit scatterers, magnetic impurities reduce $T_\text{c}$ due to
the pair-breaking term in \eqref{eq:retardedusadel}. For $L >
l_\text{sf}$ and $T\ll T_\text{c}$, as well as for $T>T_\text{c}$, the
results do not depend on the nature of the spin-flip scattering. In
general, we observe that $T_\text{c}$ strongly depends on $L$ due to
the inverse proximity effect. We also note that the difference in
damping between the normal state and the superconducting state is
small when $L\sim\xi_0$ since only a small gap develops.

The experiments of Ref.~\onlinecite{bell:cond-mat/0702461} probed the
regimes $L\ll \xi_0$ as well as $L \gg \xi_0$. We also present results
for arbitrary $L/\xi_0$. In the normal state, $G$ decreases with
increasing $L$ due to increasing bulk spin transport resistance, which
limits relaxation in R, until $L$ reaches the value of $l_{\rm sf}$
where R becomes irrelevant (inset Fig.~\ref{fig:gilbertsf}).  When
$T\ll T_\text{c}$, on the other hand, the relevant length scale for
spin penetration into S is $\xi_0$. This explains the more rapid decay
of $G-G_0$ as a function of $L$ in the superconducting state. When
$L>\xi_0$, the spin-current absorption is completely determined by the
inverse proximity effect: Spin dissipation in R by transport through S
is suppressed by the superconducting gap, and, furthermore, spin
relaxation deep in S is suppressed by the superconductivity. However,
the inverse proximity effect enhances the density of states at low
energy as well as spin-flip scattering rates close to the F$\vert$S
interface.

When $L<l_\text{sf}$, the results depend strongly on the S$\vert$R
contact described by $g$. In the right panel of
Fig.~\ref{fig:gilbertsf}, we show the temperature dependence of
$G-G_0$ for $L=900$~nm in an F$\vert$S system (no R or $g=0$). At
$T>T_\text{c}$, the damping is much smaller in the F$\vert$S system
(the right panel) than in the F$\vert$S$\vert$R system with the same
$L$ (the middle pair of curves in the left panel).  $T_\text{c}$ is
also higher since there is no inverse proximity effect at $x=L$. At
very low temperatures, $T\ll T_\text{c}$, $G-G_0$ saturates at the
same value for the F$\vert$S system as the F$\vert$S$\vert$R system
with the larger thickness, $L=1300$~nm. For such thick S, $T_\text{c}$
is unaffected by R and spins cannot diffuse through S and dissipate in
R, so that the resulting damping is the same as in the F$\vert$S
system. We also see from the right panel of Fig.~\ref{fig:gilbertsf}
that when $T\lesssim T_\text{c}$ the enhanced Gilbert damping can be
somewhat larger than above $T_\text{c}$ when spin-flip is induced by
magnetic impurities, because the induced spin accumulation of
quasiparticles with energy $k_\text{B}T\gtrsim \Delta$ experiences an
enhanced spin-flip rate through $\alpha_\text{TSTS}^\text{m}$. In the
F$\vert$S$\vert$R system, this effect is overwhelmed by the spin
accumulation drain in R.

In conclusion, our theory quantitatively reproduces the measured FMR
linewidth broadening in ferromagnet$\vert$superconductor
structures. We make additional predictions for varying system sizes
and temperatures, and the nature and strength of spin-flip
scattering. We hope to stimulate more experiments that should reveal
information about the strong inverse proximity effect and energy
dependence of spin flip scattering in these systems.

We would like to thank C. Bell and J. Aarts for discussions. This
work has been supported by NanoNed, the EC Contracts NMP-505587-1
"SFINX" and IST-033749 "DynaMax", the DFG through SFB 513 and the
Landesstiftung Baden-W\"{u}rttemberg.
  
\bibliography{/home/gudrun/janpette/artikkel/fs.bib}

\end{document}